# Anomalous thermal conductivity of $NaV_2O_5$ as compared to conventional spin-Peierls system $CuGeO_3$

A.N. Vasil'ev and V.V. Pryadun

*Low Temperature Physics Department, Moscow State University, 119899*
*Moscow,*
*Russia*

D.I. Khomskii

*Solid State Physics Department, University of Groningen, 9747 AG*
*Groningen,*
*The Netherlands*

G. Dhalenne and A. Revcolevschi

*Laboratoire de Chimie des Solides, Universite Paris-Sud, 91405 Orsay*
*Cedex,*
*France*

M. Isobe and Y. Ueda

*Institute for Solid State Physics, The University of Tokyo, 7-22-1*
*Roppongi,*
*Minato-ku, Tokyo 106, Japan*

A huge increase of thermal conductivity $\kappa$ is observed at the phase transition in stoichiometric $NaV_2O_5$. This anomaly decreases and gradually disappears with deviation from stoichiometry in $Na_{1-x}V_2O_5$ ($x$ = 0.01, 0.02, 0.03, and 0.04). This behavior is compared with that of pure and Zn-doped $CuGeO_3$ where only modest kinks in the $\kappa(T)$ curves are observed at the spin-Peierls transition. The change of $\kappa$ at critical temperature $T_c$ could be partially attributed to the opening of an energy gap $\Delta$ in the magnetic excitation spectrum excluding the scattering of thermal phonons on spin fluctuations. However, the reason for such a strong anomaly in the $\kappa(T)$ may lie not only in the different energy scales of $CuGeO_3$ and $NaV_2O_5$, but also in the different character of the phase transition in $NaV_2O_5$ which can have largely a structural origin, *e.g.* connected with the charge ordering.

75.30.Kz, 66.70.+f

June 12, 1998

Since the discovery of a spin-Peierls (SP) transition in $CuGeO_3$ [1] and $NaV_2O_5$ [2] numerous experimental and theoretical studies were continued to reveal the main features of this phenomenon in inorganic compounds. While considering that the study of $CuGeO_3$ remains very active, the main physical properties of this compound are quite well known by now. $CuGeO_3$ belongs to the orthorhombic space group Pbmm. The structure contains magnetic chains of $Cu^{2+}$ ($3d^9$, S = 1/2) ions parallel to the $c$ axis. The SP transition observed at $T_c \simeq 14$ K is marked by an exponential drop of the magnetic susceptibility and a simultaneous lattice dimerization along the $a$ and $c$ axes. Due to the magnetoelastic coupling the exchange constant between neighboring spins on the chain also alternates and the singlet ground state in the dimerized phase is separated from the band of excited triplet states by an energy gap $\Delta$ of about 25 K [3, 4].

In comparison with $CuGeO_3$, only few and somewhat contradictory data are presently available on the specific features of the phase transition in the mixed ($V^{4+}/V^{5+}$) or intermediate ($V^{4.5+}$) valence compound $NaV_2O_5$. A rapid decrease of the magnetic susceptibility below $T_c \simeq 35$ K suggests the formation of a nonmagnetic spin-singlet ground state. The simultaneous formation at this temperature of a superlattice structure characteristic of a SP transition was confirmed by X-ray measurements [5], while the opening of an energy gap ($\Delta \simeq 100$ K) in the magnetic excitation spectrum was observed in inelastic neutron experiments [6]. These data, as well as the results of NMR [7] and ESR [8] studies, are strong evidence of the presence of S = 1/2 chains considered as a necessary condition for a SP transition. On the other hand, the two-step phase transition found in specific heat and thermal expansion measurements [9] make a simple SP scenario for this transition in $NaV_2O_5$ not so evident.

The main controversy concerning the nature of the phase transition in $NaV_2O_5$ arises from the uncertainty of its crystal structure identification at room temperature. The X-ray diffraction pattern of single crystals fits well enough with both space groups $P2_1mn$ and Pmmn [10-13]. In both cases the layers of edge/corner-sharing tetragonal $VO_5$ pyramids connected in the $a - b$ plane are stacked along the $c$ axis of the structure while the $Na^{1+}$ ions are situated between these layers. In the former non-centrosymmetric structure, there exists alternating chains of nonmagnetic $V^{5+}$ ions and magnetic $V^{4+}$ ones ($3d^1$, $S = 1/2$) aligned along the $b$ axis (see e.g. Fig. 1 of Ref. [2]). In the latter structure, all the V sites are crystallographically equivalent, which results formally in a random distribution of the $V^{4+}$ and $V^{5+}$ ions, or in a localization of electrons on V-O-V molecular orbitals on the rungs of the ladders running along the $b$ axis (see e.g. Fig. 1 of Ref. [13]). In such a situation a one-dimensional magnetic behavior does not come about straightforwardly but can still be justified [13-15].

In order to clarify a picture of the phase transition in $NaV_2O_5$, more experimental data are necessary. One of the few experimental techniques not yet applied to the study of SP transitions in metal oxides is the measurement of thermal conductivity which can provide useful information on the interaction of elementary excitations in these compounds.

In this letter we report on the first, to the best of our knowledge, such measurements, which show a surprisingly strong effect in $NaV_2O_5$, in comparison with the rather weak anomaly observed in $CuGeO_3$. A possible interpretation of this effect is considered below and it is concluded that the nature of the phase transition in $NaV_2O_5$ may be more complicated than previously assumed.

Single crystals of $CuGeO_3$ and $Cu_{0.98}Zn_{0.02}GeO_3$ were grown in Paris University by a floating zone method [16]. Single crystals of stoichiometric $NaV_2O_5$ were grown in Tokyo University by the self-flux method [17]. Single crystals of non-stoichiometric $Na_{1-x}V_2O_5$ ($x = 0.01, 0.02, 0.03$, and $0.04$) were prepared by heating a small crystal of stoichiometric $NaV_2O_5$ embedded in a large quantity of $Na_{1-x}V_2O_5$ powder in an evacuated silica tube at 650°C for one week. The non-stoichiometric samples of $Na_{1-x}V_2O_5$ contain vacancies on the Na sites. Typical dimensions of the crystals used in the thermal conductivity measurements were $0.2 \times 0.8 \times 4.0$ mm$^3$. The largest dimension coincided with the dimerization axis, i.e. the $c$ axis in the case of $CuGeO_3$, and the $b$ axis in the case of $NaV_2O_5$. The smallest dimension coincided with the $a$ axis in the layered structure of $CuGeO_3$ and $c$ axis in the layered structure of $NaV_2O_5$. In most cases, measurements were performed in the direction of dimerization, but in $CuGeO_3$ thermal conductivity was measured also along the $b$ axis. The measurements were performed by the longitudinal steady-state four probe method [18] with a temperature gradient equal to 2% of the current temperature. The sample holder, surrounded by a special shell to suppress the thermal radiation, was placed into the vacuum chamber evacuated to better than $10^{-6}$ bar. The sample temperature and the temperature gradient were stabilized to better than 0.01% by an Oxford intelligent temperature controller with calibrated "Allen-Bradley" carbon resistors. Due to the relatively small dimensions of the samples the error in the determination of the absolute values of thermal conductivity was not better than 10%, while the precision on the thermal conductivity variations was about 1.5 %.

The experimental results obtained in the $c$ and $b$ directions for pure $CuGeO_3$ and for the Zn-doped sample are shown in Fig. 1. A broad phonon maximum of $\kappa$ is observed at T = 23.3 K in the $c$ direction, while it is not well pronounced in the $b$ direction. The absolute value of $\kappa$ in the $b$ direction is approximately three times lower than that in the $c$ direction. The kinks on the $\kappa(T)$ curves are evident at the temperature of the SP transition. The SP transition temperature in the Zn-doped sample is reduced to $T_c = 10.6$ K to be compared to that of the pure sample $T_c = 14.2$ K. The absolute value of $\kappa$ at the SP transition in the doped sample is significantly smaller than that in the pure sample.

The thermal conductivity of stoichiometric $NaV_2O_5$ exhibits a much more dramatic variation in the temperature range studied. As shown in Fig. 2, the thermal conductivity exhibits a broad maximum at T $\simeq$ 70 K and decreases upon approaching the SP transition. At $T_c$ a sharp upturn of $\kappa$ occurs so that a five times increase of thermal conductivity accompanied by a subsequent decrease of $\kappa$ upon cooling to liquid helium temperature takes place. A behavior similar to the one described above can be seen in non-stoichiometric samples of $Na_{1-x}V_2O_5$ ($x = 0.01, 0.02$, 0.03, and 0.04) with the progressively less pronounced maximum below $T_c$, as shown in Fig. 3. The position of the low temperature maximum remains unchanged in non-stoichiometric samples but absolute values of $\kappa$ gradually decrease with $x$. Certain variation in the high temperature slopes and the vertical shift between the curves obtained for 0 and 1% and the curves for 2, 3 and 4% Na deficiency can be tentatively ascribed to a variation in the crystals morphology. These crystals can be easily cleaved perpendicular to the $c$ axis, resulting in an uncontrolled extra scattering of the phonons on planar defects. The spin-Peierls transition temperatures in $Na_{1-x}V_2O_5$ determined from the positions of the minimum on the $\kappa(T)$ curves are shown in Fig. 4, where the solid curve is drawn as a guide for an eye. The suppression of $T_c$ by the Na-deficiency was observed earlier in X-ray critical scattering measurements [19].

Since no quantitative treatment of the transport phenomena for the spin-Peierls systems exists, we applied the Boltzmann equation for the analysis of the experimental data. In this approach the phonon thermal conductivity $\kappa \sim vlC$ is determined by the sound velocity $v$, the mean free path of phonons $l$, and the specific heat $C$ which depends on the number of elementary excitations. The interactions of the phonon and the spin subsystems can be taken into account through the variation of the phonons mean free path. In the absence of these interactions, the mean free path of the phonons increases upon cooling and eventually saturates while their number decreases. These opposite tendencies result in a phonon maximum $T_m$ on the $\kappa(T)$ curves. The estimated phonon mean free path in $CuGeO_3$ is still much less than the sample dimensions; we believe that this is caused by the phonon scattering on planar defects in the layered structure of this compound. The difference in the absolute values of the thermal conductivity in the $b$ and the $c$ directions in $CuGeO_3$ can naturally be attributed to a factor three difference in ultrasonic velocities, *i.e.* $2.7 \cdot 10^3$ $m/sec$ in the $b$ direction and $8.0 \cdot 10^3$ $m/sec$ in the $c$ direction [20]. On the other hand, the small thermal conductivity value of in the $Cu_{0.98}Zn_{0.02}GeO_3$ sample as compared with that of the pure $CuGeO_3$ sample reflects the increased number of scattering centers in the doped sample.

Anomalies observed in the temperature dependence of the thermal conductivity at the spin-Peierls transition can be naturally connected with the change of the crystal lattice and with the formation of the energy gap in the magnetic excitation spectrum. The mutual positions of the temperatures of the phonon maximum $T_m$ and of the spin-Peierls transition $T_c$ is of crucial importance for the analysis of these anomalies. The phonon maximum in $CuGeO_3$ is reached at $T_m > T_c$. In this case a possible decrease of the phonon scattering on spin fluctuation at $T_c$ cannot significantly influence the thermal conductivity, because the phonon mean free path is already large, and only modest kinks on the $\kappa(T)$ are observed. From the behavior of $NaV_2O_5$ we have to conclude that the situation here is exactly opposite. If the maximum of $\kappa$ at 70 K would have the same origin, we would not be able to explain such a strong increase of $\kappa$ below $T_c$. Therefore we have to assume that the decrease of $\kappa$ below 70 K in $NaV_2O_5$ is of a different nature, and most probably can be explained by the enhancement of phonon scattering on approaching the phase transition. The ordering occurring below $T_c$ apparently switches off this extra scattering and leads to a huge increase of the thermal conductivity. No variation of the sound velocity at $T_c$ can explain the observed behavior of the thermal conductivity in $NaV_2O_5$. While no data for the absolute value of the sound velocity are available, a longitudinal velocity variation $\Delta v/v \sim 10^{-3}$ was observed at $T_c$ preceded by a precursor effects observable below 70 K [21].

Several factors may play a role in the much stronger enhancement of the thermal conductivity below $T_c$ in $NaV_2O_5$ as compared to $CuGeO_3$. The spin gap is much larger in $NaV_2O_5$ ($\sim$100 K vs $\sim$25 K in $CuGeO_3$). Opening of this gap switches off spin-phonon scattering for phonons with frequencies less that $\Delta$. A four times larger spin gap in $NaV_2O_5$ implies that much more phonons will now strongly contribute to thermal conductivity below $T_c$. This picture could, in principle, have allowed also to explain the behavior of the thermal conductivity of $NaV_2O_5$ as a function of the deviation from stoichiometry. Magnetic measurements show that, in contrast with the relatively small suppression of $T_c$ with $x$, the drop of magnetic susceptibility below $T_c$ is strongly reduced and for $x = 0.03 - 0.04$ it nearly vanishes. This shows that the spin gap is now, to a large extent, filled by magnetic excitations. Correspondingly, the phonon-spin fluctuation scattering is being restored, which suppresses the increase of $\kappa$ below $T_c$ in non-stoichiometric samples.

However, the comparison with $CuGeO_3$ shows that pure spin-phonon scattering is hardly sufficient to explain huge increase of $\kappa$ observed. Magnetoelastic coupling in $NaV_2O_5$ is not strong enough, which follows *e.g.* from the very weak dependence of $T_c$ on magnetic field [23]. We believe that the huge increase of the thermal conductivity in $NaV_2O_5$ can be related to the specific features of the phase transition in this compound. If the crystal structure above $T_c$ is indeed that obtained in recent studies [11-13], all V ions are on the average equivalent, *i.e.* the $V^{4.5+}$ ions. Such a state can be visualized as the (dynamic) random mixture of $V^{4+}$ and $V^{5+}$. If slow enough, these charge fluctuations can strongly enhance phonon scattering and reduce thermal conductivity similar to a situation in glasses. The phase transition may then be accompanied, or even caused, by the charge ordering [22] at which this extra scattering mechanism is switched off. The opening of a spin gap may then be not so much a driving force but a consequence of this ordering. The transition from the disordered glass-like phase to an ordered one could explain the strong enhancement of the thermal conductivity below $T_c$. This picture can also explain the observation [23] that the total entropy of the transition in $NaV_2O_5$ is larger than the pure spin one.

Although our measurements do not allow to draw a final conclusion about the microscopic picture of the changes occurring at the phase transition in $NaV_2O_5$, the qualitative difference observed in the behavior of $CuGeO_3$ and $NaV_2O_5$ can be taken as an evidence that the nature of the transition in $NaV_2O_5$ is more complicated than in $CuGeO_3$ and most probably directly involves lattice and charge degrees of freedom, *e.g.* in a form of charge ordering. Further experiments, especially neutron scattering, should shed light on the microscopic picture of the spin-Peierls transition in $NaV_2O_5$ and should clarify the nature of the strong anomaly in the thermal conductivity of $NaV_2O_5$ reported in this letter.

This work was supported by the Russian Foundation for the Basic Research (Grant-in-Aid No. 96-02-19474) and by the Netherlands Foundation for the Fundamental Study of Matter (FOM). The authors are grateful to B. Buechner, D. van der Marel, A. Damascelli and M. Mostovoy for helpful discussions.

[1] M. Hase, I. Terasaki, and K. Uchinokura, Phys. Rev. Lett. **70**, 3651 (1993).

[2] M. Isobe and Y. Ueda, J. Phys. Soc. Jpn. **65**, 1178 (1996).
[3] M. Nishi, O. Fujita, and J. Akimitsu, Phys. Rev. B **50**, 6508 (1994).
[4] L.P. Regnault *et al.*, Phys. Rev. B **53**, 5579 (1996).
[5] Y. Fujii *et al.*, J. Phys. Soc. Jpn. **66**, 326 (1997).
[6] T. Yoshihama *et al.*, Physica B **234-236**, 539 (1997).
[7] T. Ohama *et al.*, J. Phys. Soc. Jpn. **66**, 545 (1997).
[8] A.N. Vasil'ev *et al.*, Phys. Rev. B **56**, 5065 (1997).
[9] M. Koeppen *et al.*, Cond-mat/9710183.
[10] A. Carpy and J. Galy, Acta Cryst. B **31**, 1481 (1975)
[11] H.-G. von Schnering *et al.*, to be published.
[12] A. Meetsma *et al.*, Submitted to Acta Cryst.
[13] H. Smolinski *et al.*, Cond-mat/9801276.
[14] A. Damascelli *et al.*, to be published.
[15] P. Horsch and F. Mack, Cond-mat/9801316.
[16] A. Revcolevschi and G. Dhalenne, Adv. Mater. **5**, 657 (1993).
[17] M. Isobe, C. Kagami, and Y. Ueda, J. Cryst. Growth **181**, 314 (1997).
[18] R. Berman, *Thermal conductivity of solids*. Clarendon Press, Oxford, 1976.
[19] H. Nakao *et al.* in Proc. of Int. Conf. on Neutron Scattering, Toronto, 1997.
[20] M. Poirier *et al.*, Phys. Rev. B **51**, 6147 (1995).
[21] P. Fertey *et al.*, submitted to Phys. Rev. B.
[22] After submission of this paper the results of NMR studies of $NaV_2O_5$ became available (T.Ohama *et al*, Submitted to Phys. Rev. Lett.), which show that there appear inequivalent V ions at T< $T_c$, so that most probably indeed a charge ordering occurs at $T_c$ in $NaV_2O_5$.
[23] B. Buechner, private communication.

FIG. 1. Temperature dependences of thermal conductivity in pure and Zn-doped $CuGeO_3$.

FIG. 2. Temperature dependences of thermal conductivity $\kappa$ and of magnetic susceptibility $\chi$ in stoichiometric $NaV_2O_5$.

FIG. 3. Temperature dependences of thermal conductivity in $Na_{1-x}V_2O_5$ ($x = 0, 0.01, 0.02, 0.03$, and $0.04$).

FIG. 4. Critical temperatures of the spin-Peierls transition in $Na_{1-x}V_2O_5$ as obtained from the thermal conductivity data. The solid curve is drawn as a guide for an eye.

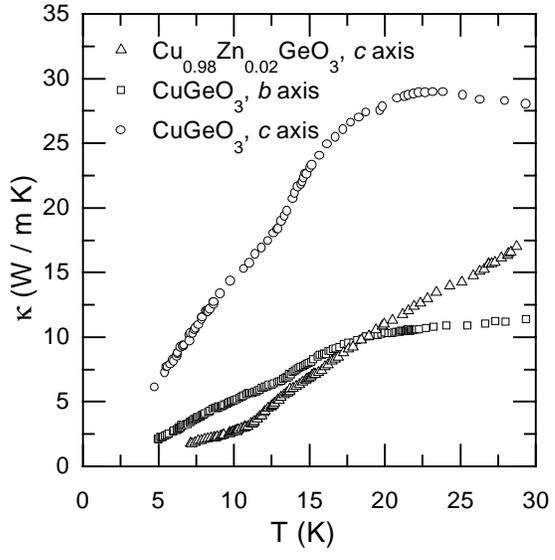

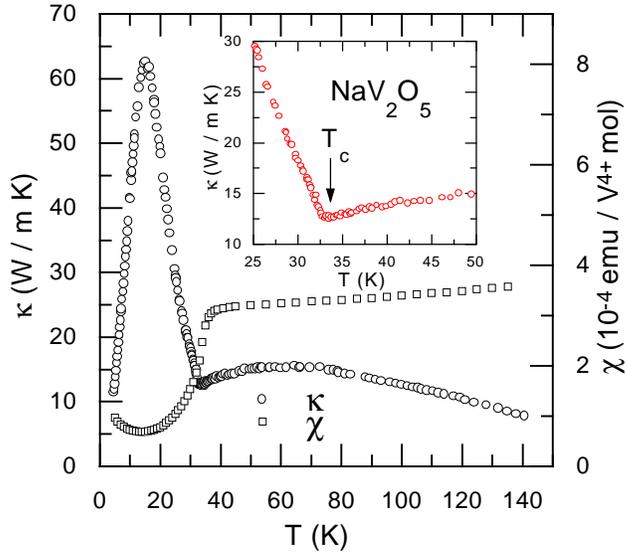

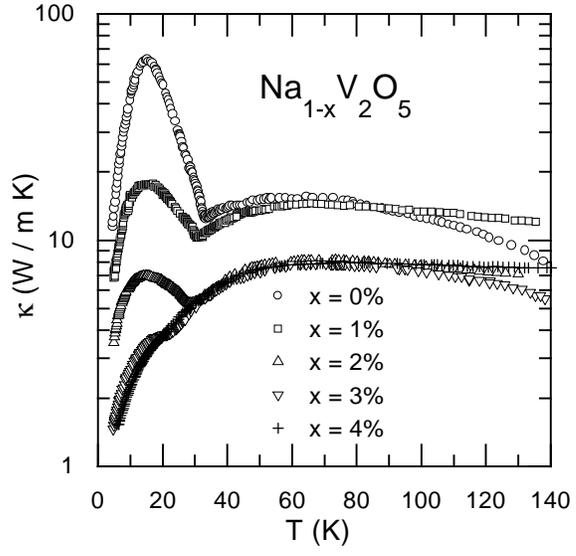

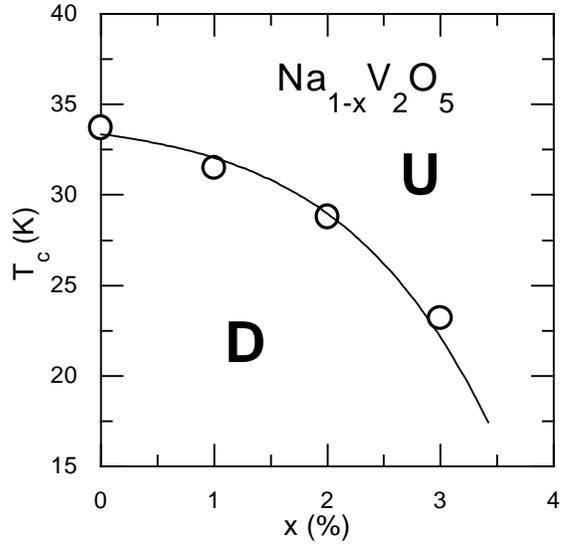